\documentstyle[12pt,aaspp]{article}
\def\etal{{\it~et\,al. }}
\def\eg{e.g.,~}
\def\kms{{\rm\,km\,s^{-1}}}

\begin{document}

\title{The Lensing Galaxy in MG1549+3047}

\centerline{\it Submitted to AJ: 1995.09.22, revised: 1995.11.20,
                accepted: 1996.01.17\,.}

\author{J.~Leh\'ar\altaffilmark{1},
        A.J.~Cooke\altaffilmark{2}}
\affil{Institute of Astronomy, Cambridge University, CB3 0HA, UK}

\author{C.R.~Lawrence}
\affil{Jet Propulsion Laboratory 169-506, Pasadena, CA 91109}

\author{A.D.~Silber}
\affil{Univ. of Washington, Dept of Astronomy FM-20, Seattle, WA 98195}

\and 
\author{G.I.~Langston}
\affil{National Radio Astronomy Observatory, Green Bank, WV 24944}

\altaffiltext{1}{Present Address: Center for Astrophysics,
      60 Garden St, Cambridge, MA 02138, jlehar@cfa.harvard.edu}
\altaffiltext{2}{Present Address: Institute for Astronomy, 
                 Royal Observatory, Edinburgh, UK}

\begin{abstract}
We have measured a velocity dispersion
for the foreground galaxy in this gravitationally lensed system.
Our dispersion confirms the prediction from lens models,
provided that the source is distant enough ($z_{_S}>0.2$).
Current interpretations of lensing statistics depend
sensitively on how the optical and mass dispersions are related. 
For $z_{_S}>0.5$, our observations favor 
$\sigma_{mass}/\sigma_{opt}\simeq1$,
but our uncertainties prevent us from 
clearly distinguishing between dynamical models. 
We could not obtain a unique mass profile from lensing constraints,
but models with a constant mass-to-light ratio are possible. 
We also found unusual rotation in the lensing galaxy,
and confirmed its redshift. 
\end{abstract}


\clearpage

\section{Introduction} 
 
Early examples of gravitational lensing (\eg\cite{wal79})
consisted of several compact images of quasars. 
When the background source is extended (\eg\cite{hew88}),
the images can join in a ring of emission around the lensing galaxy,
and multiply-imaged source features can provide
many constraints on the lensing mass distribution.
For MG1549+3047, the background source is
an extended radio lobe (\cite{leh93}), 
whose unusual structure is easily explained by a simple lensing model. 
Here we will be mostly concerned with the lensing galaxy, 
G1 (see Fig.~1 in \cite{leh93}),
which lies directly in front of the Northwest radio lobe.

Assuming an isothermal mass profile for G1, 
Leh\'ar\etal(1993) 
predicted its central velocity dispersion 
to be $\sigma_{mass}\sim230\kms$. 
A measured optical velocity dispersion $\sigma_{opt}$ 
should be comparable to this value.  
Of course, neither the mass nor the stellar profiles need be isothermal. 
For a Hubble profile stellar population 
in an isothermal gravitational potential,
$\sigma_{mass}/\sigma_{opt}=\sqrt{3/2}$ (\cite{got77}). 
This factor strongly affects the interpretation
of lensing statistics (\eg\cite{tur84}). 
However, considering a wider range of dynamical models,
and taking observational limitations into account,
Kochanek (1993) 
expects that $\sigma_{mass}/\sigma_{opt}\simeq1$ in most cases. 
Systems like MG1549+3047 provide a rare opportunity to test this prediction,
since gravitational lensing can probe the total mass distribution.


\section{Observations} 
 
We obtained long slit spectra of G1 
on the night of 1992.04.28 
at the 4.2\,m William Herschel Telescope,
using the red arm of the ISIS spectrograph. 
An EEV CCD detector was illuminated via the
R600R grating, centered on $5750{\rm\,\AA}$,
with a $5300{\rm\,\AA}$ dichroic. 
This arrangement gave $0.73{\rm\,\AA}$ per pixel, or $38\kms$,
and a spectral coverage of $5320-6180{\rm\,\AA}$.
The slit width was $1\arcsec$ on the sky ($\sim3$ pixels or $114\kms$),
and was aligned with the major axis of G1 (PA = $-151^\circ$).
The field of view along the slit was $\sim1\arcmin$.
Six 30-minute exposures were taken.
Using the same instrument configuration on 1992.05.01,
we observed one radial velocity standard star 
of spectral type K2 (K2.1=HD155642),
and two of type K5 (K5.1=HD155581, K5.2=HD156649).
The spectra were reduced using standard IRAF procedures,
and re-sampled into $44.6\kms$ bins.  
Figure~1 shows the sum of all six G1 spectra (G1.sum),
along with a standard star spectrum. 

We measured the velocity dispersion in G1 using a
standard cross-correlation method (\cite{ton79}; IRAF task ``fxcor'').
The three radial velocity standard star spectra were used as templates.
We removed 5th order spline fit baselines from the spectra
and apodized by a 10\% cosine bell, 
before Fourier transforming and 
applying a high-pass filter of $\sim10$ in wavenumber. 
Finally, we fitted a Gaussian to the maximum correlation peak
to determine its width and offset. 
We calibrated the correlation peak width using synthetic galaxy spectra
(the template star spectra convolved with 
known Gaussian dispersions, see Figure~2).
The results are summarized in Table~1\,. 
The K2.1 template gave higher cross-correlations than the K5 stars, 
so it is probably a better model of the G1 absorption features. 
We estimated the ``random'' contribution to the measurement uncertainty
by performing separate cross-correlations for each of the six G1 exposures,
and taking the rms of the dispersions. 
The ``systematic'' uncertainty, due to template mismatch, 
was estimated from the range of dispersions obtained
when different stellar templates were used. 
Combining these estimates in quadrature, 
we found $\sigma_{opt}=227\pm18\kms$.  

Our cross-correlation analysis also yields 
a radial velocity of $cz=33534\pm103\kms$ for G1. 
This result confirms a previously reported
redshift of $0.111$ (\cite{pat89}). 

Since G1 is extended, we extracted spectra for seven $1\farcs7$ apertures
parallel to the galaxy trace in the co-added data.
The cross-correlation results are given in Table~1, 
and Figure~3 shows the velocity profiles. 
The radial velocity shows a strong linear trend, 
which indicates that G1 is rotating.

\section{Discussion} 

Our velocity dispersion, $\sigma_{opt}=227\pm18\kms$,
is very close to the $\sigma_{mass}$ derived for 
an isothermal lens model (\cite{leh93}).
The two mass estimates remain comparable 
for any source redshift $z_{_S}>0.2$,
confirming the basic lensing prediction.
The radio properties and optical color of the source 
suggest that $z_{_S}>0.3$ (\cite{leh93}).
Furthermore, MG1549+3047 is a steep spectrum source
drawn from a flux-limited radio source list (\cite{leh91}).
Assuming a spectral index of $-1$, 
the selection cutoff corresponds to the flux limit
of the Parkes 2.7\,GHz radio survey (\cite{dun90}),
for which $\langle z\rangle\sim0.7$ has been measured.
Thus $z_{_S}>0.5$ is most likely.

The systematic shift in radial velocity implies that G1 is rotating. 
The velocity is still increasing in the outermost apertures,
so the maximum $v_{max}$ must exceed $200\kms$.  
G1 can be compared to an oblate isotropic rotator using the
$(v_{max}/\sigma)^*$ factor (\cite{kor82}), which exceeds 2\,.
{}From its color and radial profile, 
G1 appears to be an early-type galaxy (\cite{leh93}).
Most ellipticals rotate more slowly, 
and this level of rotation is more characteristic of S0 galaxies
or the bulges in disk galaxies (\cite{dav83}).  

The optical velocity dispersion can be compared with 
the total mass dispersion from lens models. 
Assuming a source redshift of $z_{_S}=1$,
the lens model gives an isothermal velocity dispersion 
$\sigma_{mass}=230\pm11\kms$ (\cite{leh93}),
which is the same as $\sigma_{opt}$.
If the stellar tracers are assumed to have a Hubble profile,
the expected optical dispersion will be $188\pm9\kms$ (\cite{got77});
and the strong rotation in G1 does not affect this. 
Our $\sigma_{opt}$ differs from this prediction
by almost two standard deviations,
and agrees with the expectation of Kochanek (1993). 
However, $\sigma_{mass}$ increases by $25\%$ 
if the source redshift $z_{_S}=0.3$\,.
So although our observation favors $\sigma_{mass}/\sigma_{opt}\simeq1$,
we cannot clearly distinguish between the two dynamical models.

We can also probe the dark matter distribution
by comparing the lens models with the optical surface brightness profile.
Using the IRAF task ``ellipse'',
we extracted profiles from each of the CCD images
reported in Leh\'ar\etal(1993). 
Figure~4 shows the average profile for the $I$~filter exposures.
For sky surface brightnesses $\Sigma(\theta)$, we used a model of the form:
$$
\Sigma(\theta) = \Sigma_{_0}~\left[
                 1 + (\case{\theta}{\theta_c})^2
                 \right]^{-P/2}~~,
$$
where $P=1$ corresponds to an isothermal distribution,
and $P=2$ yields a Hubble profile.
We found best fit models for each of the observed profiles,
varying $\Sigma_{_0}$, $\theta_c$, and $P$.
The best fit shape parameters,
$\theta_c=0.24\pm0.05$ and $P=1.94\pm0.05$,
did not vary with color.  
Here, the uncertainties are the rms of the parameter values,
obtained from separate fits to each profile. 
For the mass distribution, 
we used a surface density model of the same form,
with a quadrupole perturbation to account for 
the ellipticity $\epsilon=(1-b/a)$ (\cite{mir93}).
The same image constraints and search method were applied
as in Leh\'ar\etal(1993).  
There is a strong correlation between 
$\theta_c$, $\epsilon$ and $P$ (see Figure~5).
The $\chi^2$ values are very small, 
probably because the position uncertainties
in Leh\'ar\etal(1993) 
are overestimated.
To estimate the parameter uncertainties, 
we performed Monte Carlo simulations,
where the image positions were perturbed 
within their assumed $0\farcs1$ uncertainties.
When $P$ is close to its optical value, 
the best fit $\theta_c$ is also consistent with the optical profile,
provided that $\epsilon\simeq0.3$\,.
So although we cannot uniquely determine $P$,
the lens constraints are consistent with the
mass and light distributions being the same.

We are now very close to a fully self-consistent lensing model
for this system, and MG1549+3047 is promising for mass distribution studies.
The background radio lobe has many structures, 
and it straddles the multiple-imaging boundary (see \cite{leh93}).
These properties are best exploited by 
the ``Lensclean'' method (\cite{koc93b}), 
which uses all of the source structure to constrain a lens model.
The distance to the source is still unknown, however, 
and this deficiency affects many model predictions.
Obtaining the source redshift remains an important objective.

\clearpage

\acknowledgments
\noindent
We thank many people for helpful discussions, 
in particular Paul Hewett, Jordi Miralda-Escud\'e, 
Dave Carter, Chris Kochanek, Emilio Falco, and Hans Rix. 
IRAF is distributed by NOAO
which is operated by AURA Inc. under contract to the NSF. 
JL gratefully acknowleges support from a PPARC associateship,
and from NSF grant AST93-03527. 




\begin{figure}
\newpage
\begin{picture}(367,238)
\put(0,0){\includegraphics{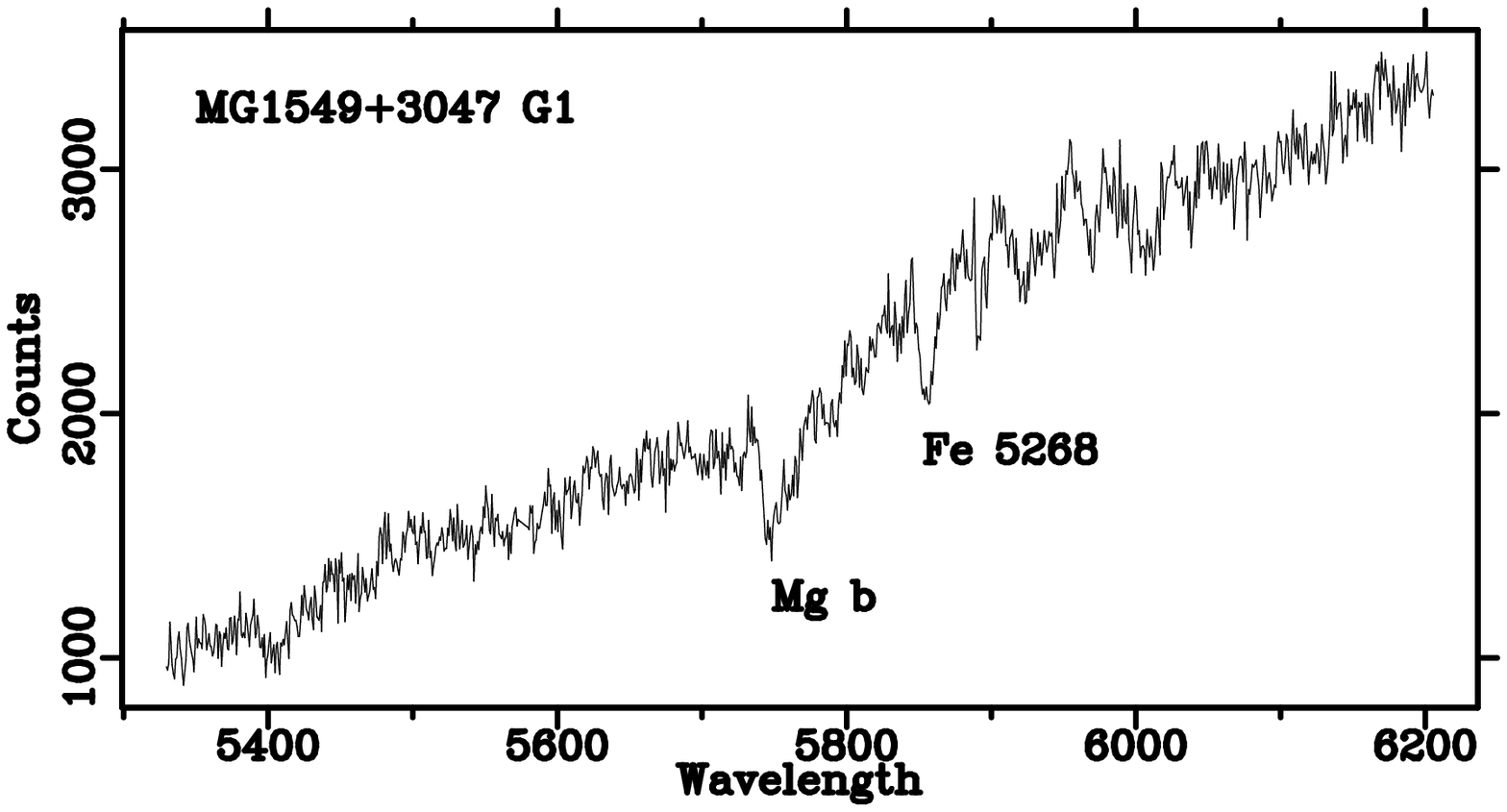}}
\end{picture}
\begin{picture}(367,238)
\put(0,0){\includegraphics{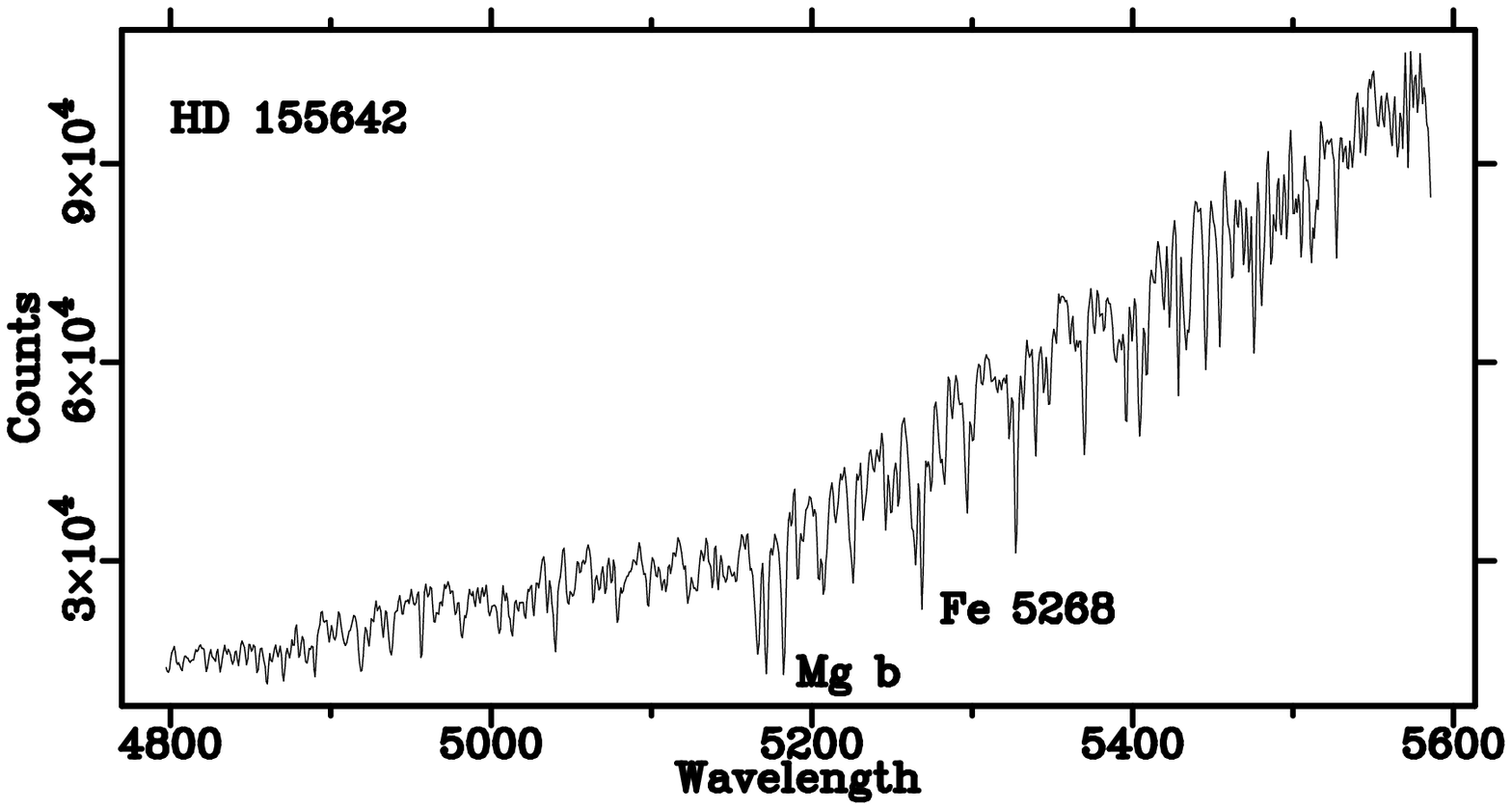}}
\end{picture}
\\
\caption{
Optical spectra for G1 (G1.sum) 
and a template star HD\,155642 (K2.1).
The G1 spectrum is the sum from all six exposures.
Wavelengths are given in Angstroms,
and two major absorption features are labelled.
}
\end{figure}

\begin{figure} 
\newpage
\begin{picture}(367,475)
\put(0,0){\includegraphics{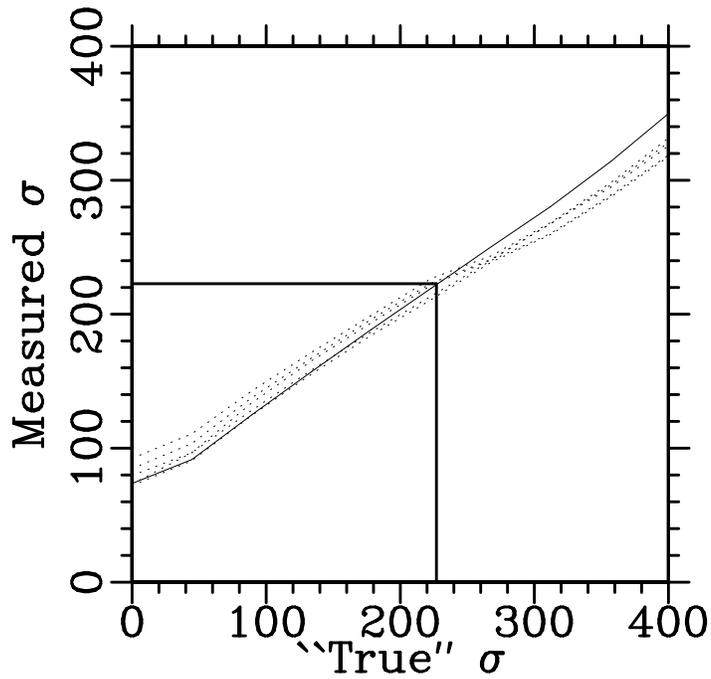}}
\end{picture}
\\
\caption{
Velocity dispersion calibration curve. 
This shows the velocity dispersion (in$\kms$)
measured from model spectra with known ``true'' dispersions.
The solid curve shows the result when K2.1 was used
for both the model and template spectrum,
and the dotted curves show other combinations of model and template. 
The heavy line shows how the observed G1 correlation width 
converts to $\sigma_{opt}$.
}
\end{figure}

\begin{figure} 
\newpage
\begin{picture}(367,475)
\put(0,0){\includegraphics{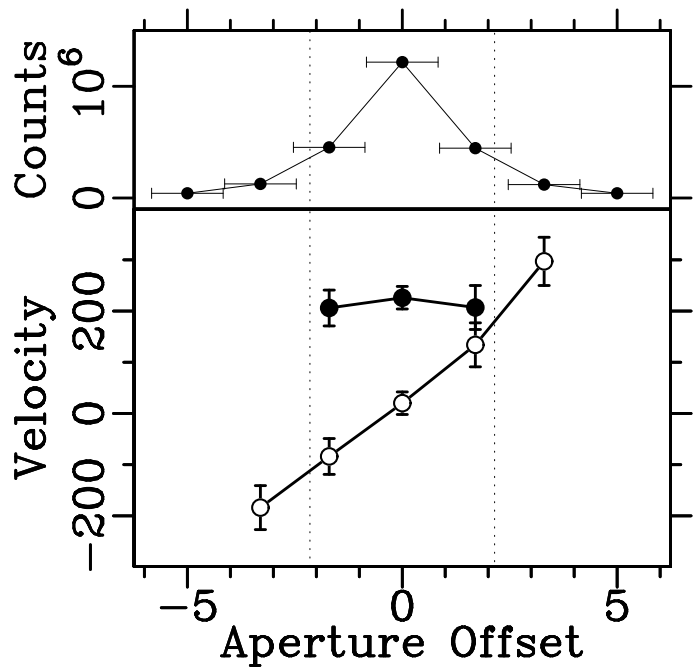}}
\end{picture}
\\
\caption{
Velocity profiles for G1, as a function of 
aperture position (in arcsec NW of G1). 
The lower panel shows the radial velocity (open circles)
and the velocity dispersion (solid dots),
both in $\kms$. 
Error bars are the formal uncertainties from ``fxcor''. 
The upper panel shows the total counts in each 
aperture, and the extent of each aperture 
is shown as a horizontal bar. 
The dotted lines in both panels show the original $4\farcs3$ aperture. 
}
\end{figure}

\begin{figure}
\begin{picture}(367,475)
\put(0,0){\includegraphics{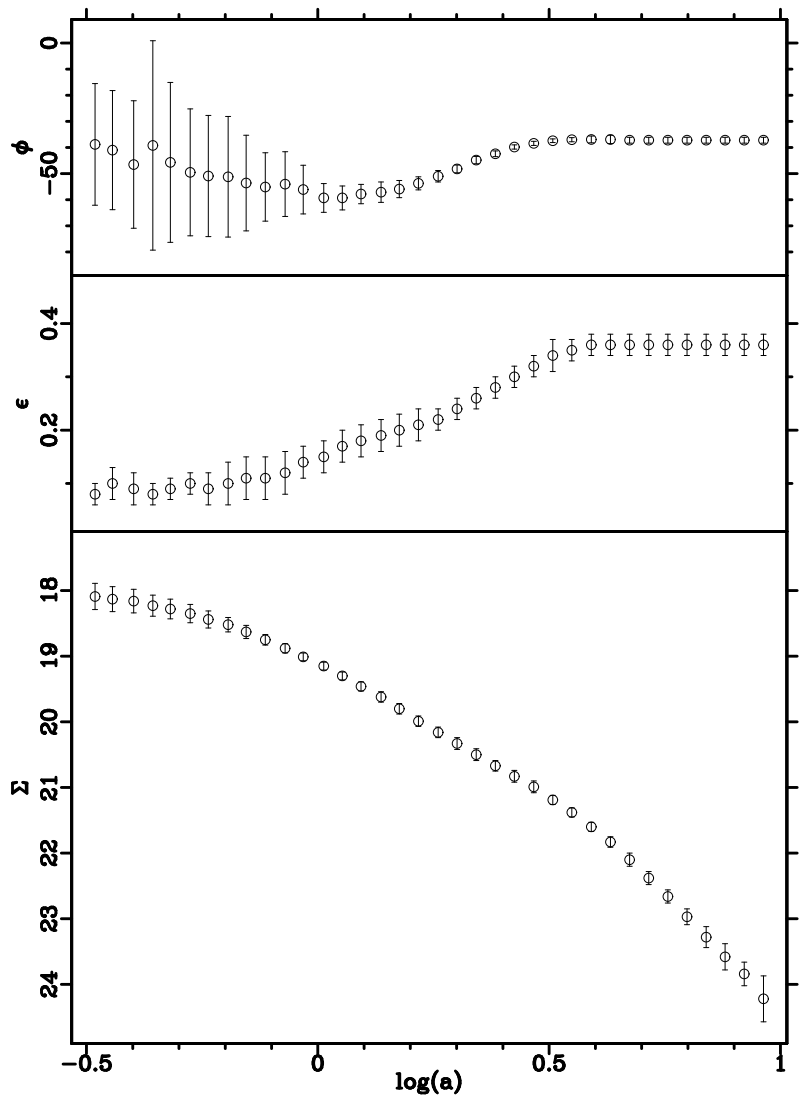}}
\end{picture}
\\
\caption{
Average surface brightness profiles for G1,
as a function of semimajor axis $a$ (in arcsec),
from the $I$~filter CCD observations of Leh\'ar\etal(1993). 
The lower panel shows the surface brightness
$\Sigma$, in ${\rm mag/arcsec^{-2}}$.  
The middle and upper panels show
the ellipticity $\epsilon=(1-b/a)$,
and the major axis orientation $\phi$
(in degrees CCW from North).
The error bars indicate the rms between profiles
from each of the separate $I$~filter exposures.
Note that both $\epsilon$ and $\phi$ vary with radius,
and that some of this may be due to atmospheric seeing. 
}
\end{figure}

\begin{figure} 
\newpage
\begin{picture}(367,475)
\put(0,0){\includegraphics{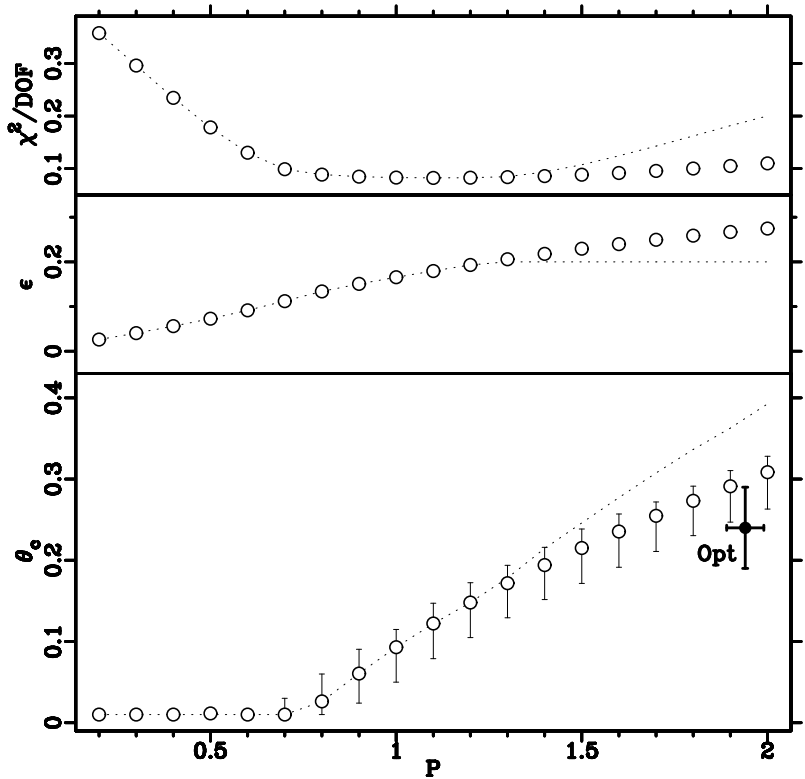}}
\end{picture}
\\
\caption{
Lens model parameters for various values of the mass index $P$.
The lower panel shows the core radius $\theta_c$ (in arcsec),
with the best fit optical values indicated.
The middle panel shows the ellipticity $\epsilon=(1-b/a)$,
and the upper panel shows the $\chi^2$ per degree of freedom.
The error bars show the 68\% confidence intervals in $\theta_c$,
from Monte-Carlo simulations (100 for each $P$).
The dotted curves show the best fit parameter values
if $\epsilon<0.2$ is applied as a constraint.
}
\end{figure}



\begin{picture}(490,633)
\put(0,0){\includegraphics{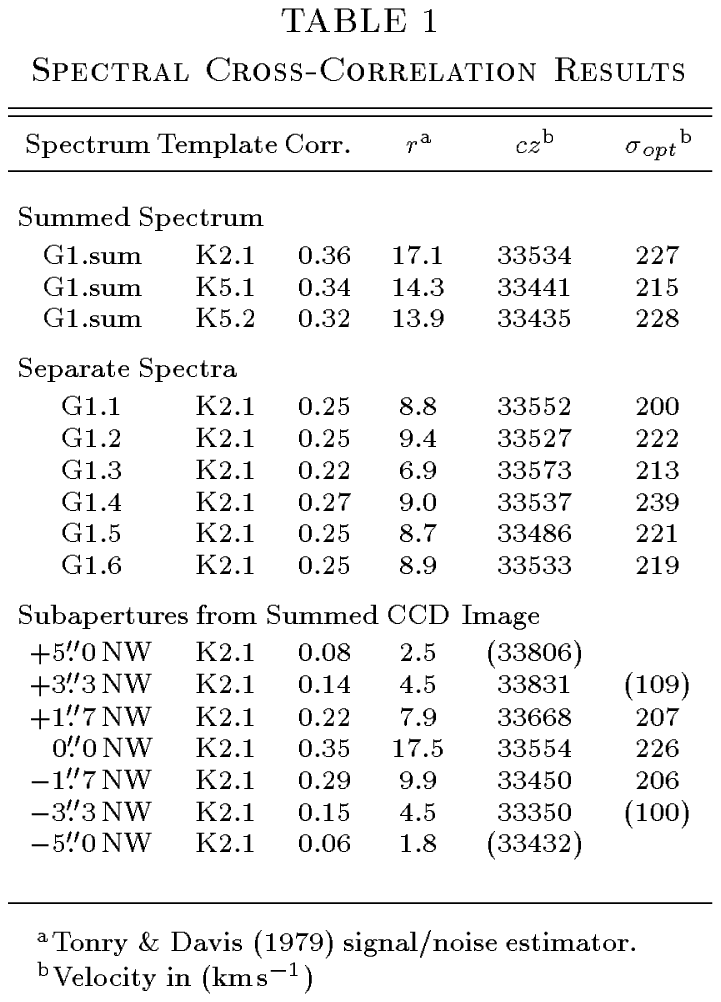}}
\end{picture}

\end{document}